\documentclass[
aps,
prl,
twocolumn,
superscriptaddress
]{revtex4-2}

\usepackage[a4paper, total={7in, 9in}]{geometry}

\usepackage{graphicx}
\usepackage{setspace}
\usepackage{color}
\usepackage{textcomp}
\usepackage{booktabs}
\usepackage[flushleft]{threeparttable}
\usepackage{comment}
\usepackage{amssymb}
\usepackage{amsmath}
\usepackage{float}
\usepackage{psfrag}
\usepackage{siunitx}
\usepackage{romannum}
\usepackage{bm}
\usepackage[colorlinks=true,linkcolor=blue,citecolor=blue,urlcolor=blue]{hyperref}
\usepackage{blindtext}
\usepackage{natbib}

\newcommand{\blue}[1]{{\textcolor{blue}{#1}}} 

\def\be{\begin{equation}}
\def\ee{\end{equation}}
\def\ba{\begin{eqnarray}}
\def\ea{\end{eqnarray}}

\begin{document}
\pagenumbering{arabic} 
\title{How Surface Viscoelasticity Eliminates Satellite Drops}

\author{Xiaocong Yang}
\affiliation{School of Astronautics, Beihang University, Beijing 100191, PR China}

\author{Qingfei Fu}
\email{fuqingfei@buaa.edu.cn}
\affiliation{School of Astronautics, Beihang University, Beijing 100191, PR China}
\affiliation{Aircraft and Propulsion Laboratory, Ningbo Institute of Technology, Beihang University, Ningbo 315800, PR China}

\author{Lijun Yang}
\email{yanglijun@buaa.edu.cn}
\affiliation{School of Astronautics, Beihang University, Beijing 100191, PR China}

\author{Bingqiang Ji}
\email{bingqiangji@buaa.edu.cn}
\affiliation{School of Astronautics, Beihang University, Beijing 100191, PR China}
\affiliation{Aircraft and Propulsion Laboratory, Ningbo Institute of Technology, Beihang University, Ningbo 315800, PR China}

\begin{abstract} 
Satellite drops form widely during the breakup of liquid threads and constitute deleterious byproducts across a broad range of industrial technologies, yet their complete suppression has remained a longstanding challenge. Here, we experimentally report that surface viscoelasticity introduced by a small amount of bovine serum albumin can fully eliminate satellite drops, via modulating the localized spatiotemporal topology near pinch-off singularity. During the late thinning stage, surface shear viscosity counterbalances the capillary and stably anchors the thinning neck at the midpoint between the two primary beads, fundamentally precluding satellite droplet formation. We further propose a scaling law to describe the unique self-similar thinning behaviors dictated by surface viscosity. These results shed light on the thinning dynamics of liquid threads with surface rheology and offer a promising strategy to eliminate satellite drops in practical applications. 
\end{abstract}

\maketitle
\newpage

Satellite drops spawned during the breakup of liquid threads are ubiquitous and serve as detrimental byproducts that degrade process fidelity and contaminate devices in high-precision applications, including microfluidics, inkjet printing, bioprinting, and extreme ultraviolet lithography \cite{Derby_2010, Chen_2025, Gupta_2025, Lyu_2024, He_2017, Rohani_2010}. Suppressing such drops is thus critical, yet remains highly challenging because of the complex dynamics involved, from the initial Rayleigh-Plateau instability, through nonlinear evolution and self-similar thinning, to the final singular pinch-off \cite{Derby_2010, Huang_2019, Eggers_2008}. Active strategies based on waveform optimization can suppress satellites but require additional hardware that increases system complexity and cost \cite{Yang_2018, Hu_2021, Zhang_2022, Lashkaripour_2024}. Passive alternatives relying on raising shear viscosity or viscoelasticity offer only a narrow operation window and risk nozzle clogging \cite{Hoath_2012, Sen_2021, Lv_2024, Zhou_2018, Eggers_2020}. This motivates the search for passive strategies that eliminate satellite drops without invoking bulk rheological modifications.

Surfactants are widely present in natural and industrial liquids \cite{Langevin_2014, Manikantan_2020}. Without altering liquid bulk properties, a minute amount of surfactant can introduce extra surface stresses that dominate at small scales and profoundly reshape interfacial flows \cite{Ponce-Torres_2020, Wee_2020, Martínez-Calvo_2020b, Erni_2011, Choi_2011, Montanero_2021}, opening a new avenue for satellite drop modulation. Small-molecule surfactants can reduce the thinning rate and modify the liquid thread topology near pinch-off via Marangoni stresses, yet they fail to eliminate satellite drops entirely \cite{McGough_2006,Craster_2009,Ponce-Torres_2017,Martínez-Calvo_2018,Martínez-Calvo_2020,Kamat_2018}. In contrast, macromolecular surfactants such as gel-like proteins form an adsorbed layer with an interconnected network that exhibits strong intrinsic surface viscoelasticity \cite{Sharma_2011,Zhong_2022}. Previous studies theoretically suggest that surface viscous stresses can overwhelm Marangoni stresses during thread thinning and effectively reduce satellite drop size, yet lacking experimental evidences \cite{Yang_2025, Ponce-Torres_2020, Wee_2020, Martínez-Calvo_2020b}. Meanwhile, surface elasticity is reported to fundamentally change the outcomes of film rupture and jetting of bubble bursting at liquid surfaces, whereas its role in liquid thread breakup is unclear \cite{Tammaro_2021,Ji_2023}. How surface viscoelasticity modulates satellite drop formation threrefore remains an open question.

In this Letter, we experimentally demonstrate that surface viscoelasticity completely kills satellite drops without appreciably altering the global flow. We further develop a theoretical model that elucidates the underlying mechanism by revealing the dominant role of surface shear viscosity, and propose a scaling law to describe the associated self-similar thinning behavior near pinch-off. These findings highlight the ability of surface rheology as a robust passive strategy for satellite drop elimination.

\begin{figure}[t!] 
\begin{center} 
\includegraphics[width=0.48\textwidth]{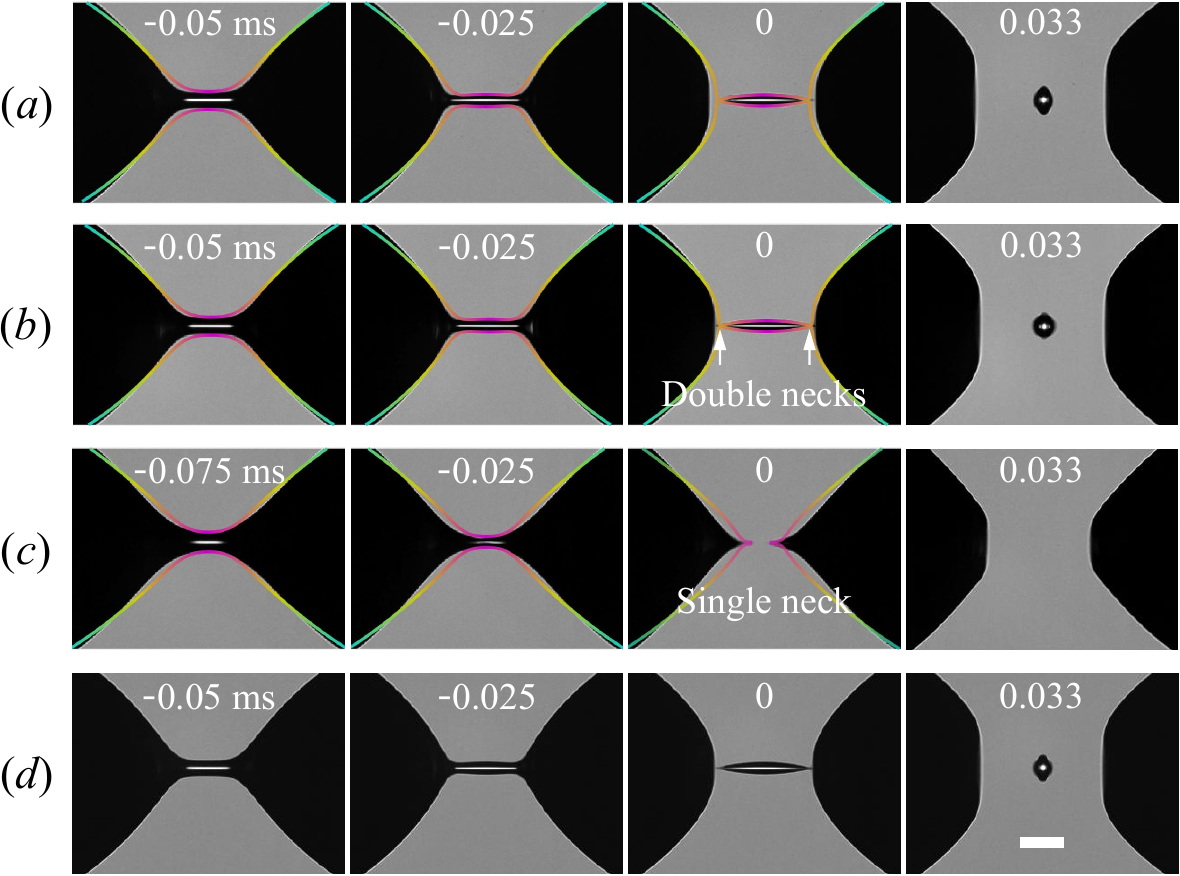}
\caption{\setstretch{1} High-speed observations of liquid thread breakup: (a) clean aqueous solution, BSA solutions with concentration $C$ of (b) 0.001 and (c) 0.1 g/L, (d) SDBS solution of $0.418\ \mathrm{g/L}$ (1 CMC). The frame at rupture is defined as $t=0$. The colored gradient curves are the thread profiles obtained numerically. Scale bar is 0.1 mm.
}
\label{fig:experiment}
\end{center}
\end{figure}

To isolate the effect of surface viscoelasticity, we adopt a widely studied globular protein, bovine serum albumin (BSA), dissolved in a NaCl buffer solution, yielding a viscoelastic liquid surface while the bulk remains Newtonian \cite{Sharma_2011}. Owing to the slow adsorption kinetics of BSA \cite{Zhong_2022,Dhar_2010}, surface viscoelasticity is tuned by varying the BSA concentration ($C=0.001-5$ g/L) and aging time ($t_a=100-600$ s, measured from fresh surface formation). A cylindrical thread of radius $R_0=0.32$ mm and length $L\approx4R_0$ is pinned between two identical blunt needles, aged for a prescribed time, and then quasi-statically thinned by slow liquid withdrawal (see Supplemental Material Sec. \blue{S1} for experimental details). A central neck progressively thins into a slender filament connecting two primary beads. At $C=0.001$ g/L [Fig. \ref{fig:experiment}(b)], the minimum radius shifts to the two filament-bead junctions, forming a double-neck structure that pinches off into a satellite drop, a behavior same to that of an inviscid clean liquid thread [Fig. \ref{fig:experiment}(a)] \cite{Castrejon-Pita_2012, Huang_2019}.

\begin{figure*}[t!] 
\begin{center} 
\includegraphics[width=0.9\textwidth]{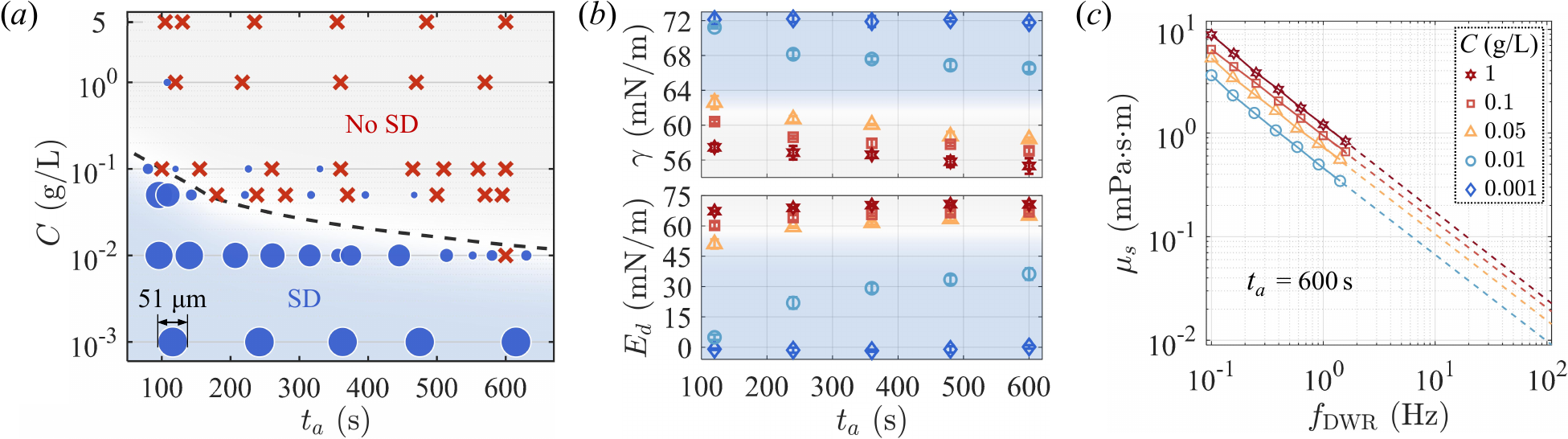}
\caption{\setstretch{1}(a) Regime map of liquid thread breakup, showing two regimes of with and without satellite drop (SD and No SD), where the size of the solid circle scales linearly with the measured satellite drop size. (b) Surface tension ($\gamma$) and surface dilatational elastic modulus ($E_d$, at an oscillating frequency of 1 Hz) of BSA solutions. (c) Surface shear viscosity ($\mu_s$) as a function of oscillatory shear frequency ($f_{\mathrm{DWR}}$) at $t_a = 600$ s, where the data at $C=0.001$ g/L are below the lower boundary. The extrapolated value at $f_{\mathrm{DWR}} = 100\ \mathrm{Hz}$ is taken as the characteristic result. }
\label{fig:combination}
\end{center}
\end{figure*}

In stark contrast, liquid thread with $C=0.1$ g/L exhibits a distinct pinch-off behavior [Fig. \ref{fig:experiment}(c)]. The neck stays centered during thinning and satellite drops vanish entirely. For a clean liquid thread, satellite drops form when the Ohnesorge number ($Oh=\mu/(\rho R_0\gamma_0)^{1/2}$, comparing viscous to inertio-capillary effects) is below 0.1 \cite{Notz_2004, Dong_2006, Detlef_2022}, where $\mu$, $\rho$, $\gamma_0$ are the viscosity, density, and initial surface tension of the liquid, respectively. Here, $\gamma_0$ is measured by the pendent drop method (Fig. \ref{fig:combination}b), and the instantaneous surface tension is described by the equation of state \cite{Martínez-Calvo_2020}
\begin{equation}
 \gamma=
 \begin{cases}
 \gamma_f - \mathcal{A} \ln\!\left(1+\Gamma / \Gamma_{m}\right), & \Gamma<\Gamma_{m},\\[6pt]
 \gamma_{\min}, & \Gamma\ge \Gamma_{m},
 \end{cases}
 \label{eq:gamma}
\end{equation}
where $\mathcal{A}=(\gamma_f-\gamma_{\min})/\ln 2$. $\gamma_f$, $\gamma_{\min}$, and $\Gamma_m$ are the surface tension of a clean surface, minimum surface tension, and the maximum adsorption concentration, respectively [Supplementary Material Sec. \blue{S3} and \blue{S4}]. Trace BSA addition leaves the bulk properties unchanged, yielding $Oh=0.006-0.008$ for all cases. The disappearance of satellite drops at such low $Oh$ therefore indicates that BSA introduces additional surface effects beyond capillarity.

To rule out the effect of Marangoni stresses, we perform control experiments with the small-molecule surfactant sodium dodecylbenzenesulfonate (SDBS). Over a wide concentration range including its critical micelle concentration (CMC), SDBS threads exhibit breakup behavior and satellite drop sizes similar to the neat case [Fig. \ref{fig:experiment}(d) and Supplemental Material Sec. \blue{S2}]. We therefore attribute satellite-drop elimination to the intrinsic surface viscoelasticity of the adsorbed BSA layer.

\begin{figure}[b!] 
\begin{center} 
\includegraphics[width=0.4\textwidth]{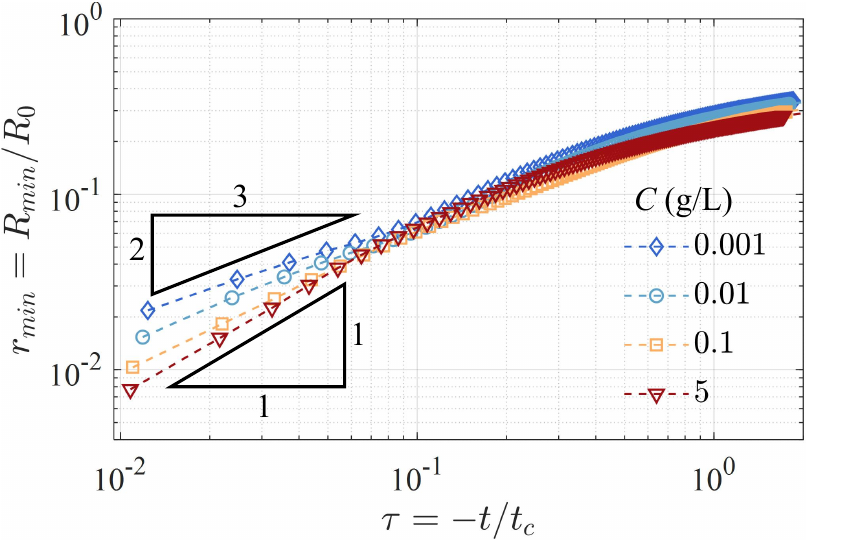}
\caption{\setstretch{1}Time variation of minimum neck radius till rupture for BSA solution threads of different concentrations. Here, $\gamma_0 \approx 55 \mathrm{mN/m}$ when $C = 5\mathrm{g/L}$.
}
\label{fig:scalinglaw}
\end{center}
\end{figure}

Interestingly, Fig. \ref{fig:experiment} shows that surface viscoelasticity barely alters the global thread morphology throughout most of the thinning process, yet modifies the local topology only in a small spatiotemporal region near the pinch-off singularity. To highlight this feature, Fig. \ref{fig:scalinglaw} plots the nondimensional neck radius $r_{min}=R_{min}/R_0$ versus $\tau=-t/t_c$, where $R_{min}$ is the radius at the thinnest position of the thread and $t_c=(\rho R_0^3/\gamma_0)^{1/2}$ is the capillary time. Indeed, the neck radius evolution remains coincident for threads with different BSA concentrations throughout most of the thinning process, and diverges only near the final pinch-off. Specifically, for threads with small $C$ ($\leq0.01$ g/L), the near-pinch-off dynamics follows $r_{min}\sim\tau^{2/3}$, characteristic of the inertio-capillary dominated self-similar regime of a clean inviscid thread \cite{Day_1998}. In contrast, for large $C$ ($\geq0.1$ g/L), the late-stage scaling deviates markedly, with the exponent increasing to approximately 1, indicating a distinct self-similar thinning regime governed by surface viscoelasticity.

To map the parameter space of satellite-drop formation, we vary the BSA concentration and aging time over broad ranges. Two distinct regimes, with and without satellite drop (SD and No SD), are separated by a clear boundary [Fig. \ref{fig:combination}(a)], with the critical concentration decreasing from 0.1 g/L at $t_a=100$ s to 0.02 g/L at $t_a=600$ s. Moreover, the satellite-drop radius $R_{sd}$ also decreases markedly with increasing $C$ and $t_a$, until satellite formation is completely suppressed.
These results shows that minute amount BSA of only $O(10-100)$ ppm can effectively eliminate the satellite drop.

In fact, surface viscoelasticity is characterized by four independent rheological parameters: surface shear and dilatational viscosities ($\mu_s$ and $\kappa_s$) and surface shear and dilatational elastic moduli ($E_s$ and $E_d$) \cite{Jaensson_2018,Erni_2011,Fuller_2012}. 
To quantify the contributions from the above four parameters, we measure the surface shear rheology using a double wall ring module mounted on a rotational shear rheometer in oscillatory mode and the surface dilatational rheology via oscillating pendant drop method \cite{Sharma_2011,Vandebril_2010, Yang_2019,Benjamins_1996} (see Supplementary Material Sec. \blue{S5-S6} for measurement detail). Results show that $E_s$ is smaller than $E_d$ by about one order of magnitude, indicating that surface elasticity is dominated by dilatational elasticity in our experiments. Meanwhile, the values of $\kappa_d$ and $\mu_s$ are close in magnitude. In this case, our previous study points out that the effect of surface viscosity on liquid thread thinning is dominated by surface shear viscosity \cite{Yang_2025, Yang_2026}. This indicates that the surface rheology modified pinch-off dynamics in our experiments may come from the effects of surface shear viscosity and dilatational elasticity.

Our surface rheology measurement shows that $E_d$ and $\mu_s$ indeed increase with $C$ and $t_a$. BSA solutions of $C \ge 0.05$ g/L exhibit an $E_d$ falling between 50 and 70 $\mathrm{mN\cdot m^{-1}}$ [Fig. \ref{fig:combination}(b)], while small-molecule surfactants exhibit an apparent elastic modulus of approximately 20 $\mathrm{mN\cdot m^{-1}}$ induced by Marangoni effect only near their CMC \cite{Kim_2017,Langevin_2014}. Fig. \ref{fig:combination}(c) further shows that $\mu_s$ decreases linearly with the shear frequency ($f_{\mathrm{DWR}}$) in log-log coordinates, as previously documented \citep{Sharma_2011}. Considering the liquid thread breakup occurs within 10 ms, we choose $f_{\mathrm{DWR}}=10^{2}\ \mathrm{Hz}$ as the characteristic shear frequency. Extrapolation gives $\mu_s \sim O(10^{-2})\ \mathrm{mPa\cdot s\cdot m}$ at $10^{2}\ \mathrm{s^{-1}}$ for BSA solutions of $C \ge 0.05$ g/L. For comparison, the reported $\mu_s$ values for small-molecule surfactants are about three orders of magnitude lower, typically below $O(10^{-5})\ \mathrm{mPa\cdot s\cdot m}$ \cite{Ponce-Torres_2020, Zell_2014}. These results indicate the  much stronger intrinsic surface viscoelasticity of macromolecular surfactants like BSA compared to small-molecule surfactants.

We then investigate the role of the above two surface rheological parameters in liquid thread pinch-off by deriving a one-dimensional theoretical model. Gravitational effects are safely ignored due to the much smaller length scale compared to capillary length. The velocity ($\mathbf{u}$) and pressure ($p$) fields are described by the continuity and Navier-Stokes equation, $\nabla\mathbf{u}=0$ and $\rho(\partial \mathbf{u}/\partial{t} + \mathbf{u} \cdot \nabla\mathbf{u}) = -\nabla{p} + \nabla \cdot \mathbf{T}$, respectively, where $\mathbf{T}=\mu [\nabla\mathbf{u}+(\nabla\mathbf{u})^{\rm{T}}]$ is the bulk viscous stress tensor. Neglecting the dynamic effects from the surrounding gas \cite{Martínez-Calvo_2020}, the dynamic boundary condition at the interface yields $\mathbf{n} \cdot (p \mathbf{I} - \mathbf{T}) + \nabla^s \cdot \bm{\tau^s} = 0$, where $\mathbf{n}$ is the unit normal vector, 
and $\nabla^s=\mathbf{I^s}\cdot \nabla$ is the surface gradient operator with $\mathbf{I^s}$ the surface identity tensor. $\bm{\tau^s} = \gamma\mathbf{I^s}+\bm{\tau_e}$ is the surface stress tensor, where $\bm{\tau_e}$ is the extra surface stress contributed by the surface viscoelasticity. For simplicity, we adopt a linear surface rheology model and let $\bm{\tau_e}=\bm{\tau_{ev}}+\bm{\tau_{ee}}$, where $\bm{\tau_{ev}}$ and $\bm{\tau_{ee}}$ respectively represent the extra surface stress from surface viscosity and elasticity. Here, $\bm{\tau_{ev}}$ is expressed based on the classical Boussinesq-Scriven approximation \cite{Boussinesq_1913, Scriven_1960},
\begin{equation}
   \bm{\tau_{ev}}=\kappa_s(\nabla^s\cdot\mathbf{u^s})\mathbf{I^s}+\mu_s[2\mathbf{U^s}-(\nabla^s\cdot\mathbf{u^s})\mathbf{I^s}],
\label{eq:Tsv}
\end{equation}
where $\mathbf{U^s}=1/2[\nabla^s \mathbf{u^s}\cdot\mathbf{I^s}+\mathbf{I^s}\cdot(\nabla^s \mathbf{u^s})^T]$ is the surface rate-of-deformation tensor and $\mathbf{u^s}$ is the surface velocity vector. Following our previous study, $\bm{\tau_{ee}}$ is described by a linear elastic model written as \cite{Jaensson_2018}
\begin{equation}
   \bm{\tau_{ee}}=E_d(\nabla^s\cdot\mathbf{v^s})\mathbf{I^s}+E_s[2\mathbf{V^s}-(\nabla^s\cdot\mathbf{v^s})\mathbf{I^s}],
\label{eq:Tse}
\end{equation}
where $\mathbf{V^s}=1/2[\nabla^s \mathbf{v^s}\cdot\mathbf{I^s}+\mathbf{I^s}\cdot(\nabla^s \mathbf{v^s})^T]$ is the surface strain tensor and $\mathbf{v^s}$ is the surface displacement vector. We set $\kappa_s = E_s = 0$ considering the negligible role of surface shear elasticity and dilatational viscosity. Following previous studies \cite{Martínez-Calvo_2020, Zhong_2022, Yang_2025}, $\mu_s$ and $E_d$ is assumed to vary linearly with BSA surface concentration $\Gamma$, i.e., $\mu_{s}=\mu_{s0}\Gamma/\Gamma_0$ and $E_{d}=E_{d0}\Gamma/\Gamma_0$, where $\mu_{s0}$ and $E_{d0}$ are the values corresponding to the initial surface concentration $\Gamma_0$. An additional equation for $\Gamma$ should be introduced. As the time scales of bulk and surface diffusion, adsorption, and desorption are at least 2 orders of magnitude smaller than the convection time scale in our experiments [Supplemental Material Sec. \blue{S7}], $\Gamma$ is dominated by surface convection and we have $\partial \Gamma/\partial{t} +\nabla^s \cdot (\Gamma \mathbf{u}) = 0$ \cite{Stone_1990_pof}.

Scaling length, time, stress, tension, and concentration by $R_0$, $t_c$, $\gamma_0/R_0$, $\gamma_0$, and $\Gamma_{m}$, respectively, yields the one-dimensional model (see Supplemental Material Sec. \blue{S7-S9} for derivation),
\begin{equation}
\begin{aligned}
  & w_{\hat{t}}+ww_z= 3 Oh (S^2 w_z)_z /S^2 + 2\hat{\gamma}_z/S \\ & - (\hat{\gamma} \zeta)_z + 2 E_{cd} \hat{\Gamma} \phi_{ed} /(S \hat{\Gamma}_0) + (\zeta Oh_{s\mu}  \hat{\Gamma} \phi_{\mu})_z / \hat{\Gamma}_0 \\ & - 2(Oh_{s\mu} \hat{\Gamma} \phi_{\mu})_z /(S \hat{\Gamma}_0) + 5Oh_{s\mu}(\hat{\Gamma} S w_z)_z / (S^2 \hat{\Gamma}_0),
\end{aligned}
\label{eq:momentum}
\end{equation}
\begin{equation}
  S_{\hat{t}}+S_z w + S w_z /2 = 0,
\label{eq:S}
\end{equation}
\begin{equation}
  \hat{\Gamma}_{\hat{t}}+\hat{\Gamma}_z w + \hat{\Gamma}(Sw_z+S_z w)/S - \zeta\hat{\Gamma}(Sw_z/2+S_z w) = 0,
\label{eq:Gamma}
\end{equation}
here $\hat{t}$, $S$, $r$, $z$, $w$, $\hat{\Gamma}$ and $\hat{\Gamma}_0$ denote nondimensional time, surface position, radial coordinate, axial coordinate, axial velocity, surface concentration, and initial surface concentration, respectively. The subscripts denote corresponding partial derivatives. $\zeta = 1/ S(1+S_z^2)^{1/2} - S_{zz}/(1+S_z^2)^{3/2}$ is the curvature. $\phi_{ed} = S_z \Delta r_{zz} + S_{zz} \Delta r_z + (S\Delta r_z - \Delta r S_z)/S^2$, $\phi_{\mu}=(Sw)_z/S - \zeta(Sw_z/2+S_z w)$, with $\Delta r = S-S_0$ represents the displacement of surface. $\hat{\gamma}=\hat{\gamma}_f-\beta\ln(1+\hat{\Gamma})$, where $\hat{\gamma}_f = \gamma_f/\gamma_0$, and the Marangoni number $\beta=\mathcal{A}/\gamma_0$ compares the Marangoni elasticity to capillary. The surface Ohnesorge number $Oh_{s\mu}=\mu_{s0}/(\rho\gamma_0R_0^3)^{1/2}$ characterizes the ratio of surface shear viscosity to inertio-capillary, while the surface elastocapillary number $E_{cd}=E_{d0}/\gamma_0$ characterizes the ratio of the surface dilatational elasticity to capillary.

The above equations are numerically solved using a finite difference method with a fifth-order WENO scheme \cite{Borges_2008} (see Supplementary Material Sec. \blue{S10} for detail), with the computation terminated when $S_{\min}<10^{-4}$.
To valid our theoretical model, the pinch-off processes of clean and BSA-coated threads in Fig. \ref{fig:experiment} are numerically solved. The numerical calculation reproduce the evolution of thread profile as well as the satellite drop elimination, demonstrating that our theoretical model can well describe the pinch-off dynamics of liquid thread with surface rheology. 

\begin{figure*}[ht!]
\begin{center} 
\includegraphics[width=0.9\textwidth]{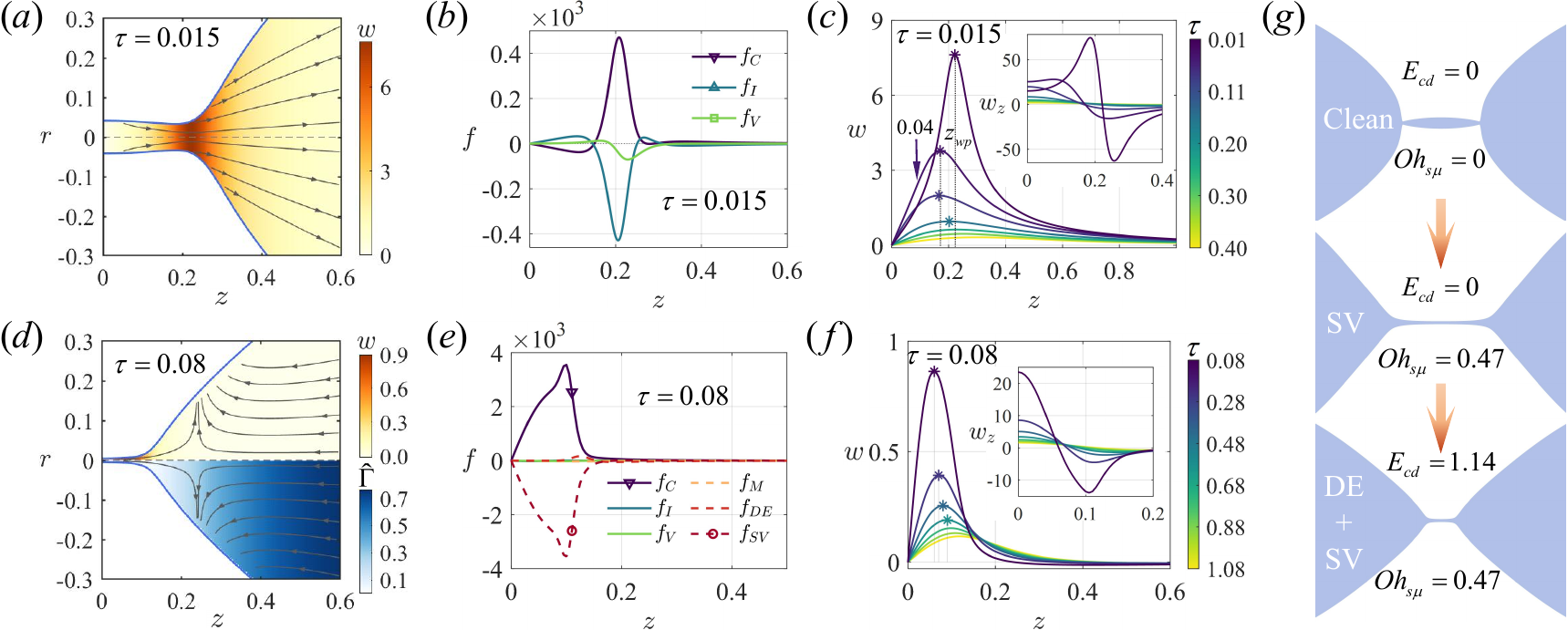}
\caption{\setstretch{1}(a) Streamlines and axial velocity ($w$) field and (b) axial distribution of forces at $\tau=0.015$, and (c) axial distribution of $w$ at different $\tau$ for a clean thread with $Oh=0.0066$. (d) Streamlines and $w$ and BSA surface concentration ($\hat{\Gamma}$) fields and (e) axial distribution of forces at $\tau=0.08$, and (f) axial distribution of $w$ at different $\tau$ for a BSA solution thread with $Oh = 0.0074$, $\hat{\Gamma}_0 = 0.68$, $Oh_{s\mu}=0.47$, $E_{cd}=1.14$, and $\beta=0.51$. Insets in (b) and (e) shows the axial distribution of $w_z$. (g) Thread Profiles of thread near pinch-off with (bottom) and without (top) surface dilatational elasticity.
}
\label{fig:mechanism}
\end{center}
\end{figure*}

To understand how surface viscoelasticity eliminates satellite drops, we first discuss their formation during the pinch-off of a clean thread at $Oh = 0.0066$ [corresponding to Fig. \ref{fig:experiment}(a)]. Force analysis shows that the balance between capillary $f_C$ and inertial [$f_I=-(w_t+ww_z)$, scaled by $\gamma_0/R_0$] forces dominates the entire thinning process, while the bulk viscous force $f_V$ plays only a minor role [Fig. \ref{fig:mechanism}(b)], consistent with an essentially inviscid flow \cite{Huang_2019, Castrejon-Pita_2012}. Capillarity drives the liquid from the central neck toward both ends. However, as the velocity $w$ and velocity gradient $w_z$ increase, the nonlinear convective term $ww_z$ becomes significant [Fig. \ref{fig:mechanism}(c)]. Note the values of $w_z$ at both sides of the peak of $w$ (at $z_{wp}$) are opposite. At $z<z_{wp}$, $ww_z>0$ and increases to overweight $f_C$ at the late stage ($\tau\le0.04$), making $w_t<0$ and $w$ decrease. In contrast, at $z>z_{wp}$, $ww_z<0$, always giving $w_t>0$ and an increasing $w$. Consequently, $z_{wp}$ migrates outwards and the peak of $w_z$ shifts outwards from the thread center. Since the thinning rate can be expressed as $dR/dt \approx -Rw_z/2$ based on Eq. \ref{eq:S}, this causes the fastest thinning position moves outwards, shifting the neck from the center ($z=0$ at $\tau\ge 0.04$) to the peak positions of $w_z$ ($z=0.23$ at $\tau=0.015$). Pinch-off then occurs at the double necks, trapping a satellite drop.

The influence of surface viscoelasticity is then revealed from the numerical results at $Oh = 0.0074$, $Oh_{s\mu}=0.47$, $E_{cd}=1.14$, and $\beta=0.51$ [corresponding to Fig. \ref{fig:experiment}(c)]. In this case, large surface concentration gradient $\hat{\Gamma}_z$ as well as $w_z$ occurs at the region with notable axial velocity $w$ at the late thinning stage [Fig. \ref{fig:mechanism}(d)]. This generates a strong surface shear viscous force, which balances capillary force and overwhelms other forces including $f_I$, $f_V$, Marangoni force $f_M$, and surface dilatational elastic force $f_{DE}$ [Fig. \ref{fig:mechanism}(e)]. The large $f_{SV}$ suppresses the effects of the nonlinear term $ww_z$ and facilitates a continuously increasing $w$ in the thread region ($z<0.23$), thereby maintaining the peak of $w_z$ as well as the fastest thinning position at the center [Fig. \ref{fig:mechanism}(f)]. This forces the neck to anchor at the thread center and completely suppresses satellite drop formation. Though smaller in comparison, cutting off the surface dilatational elasticity effect by setting $E_{cd}=0$ generates a long thin thread at the late stage [Fig. \ref{fig:mechanism}(g)], a behavior reported also by previous numerical studies \cite{Martínez-Calvo_2020b,Yang_2025}. Actually, surface dilatational elastic stress resists surface deformation and thus shortens the length of liquid thread, further inhibiting the formation of new necks beyond the thread center and promoting the effect of surface shear viscosity in satellite drop elimination.

Finally, we derive a scaling law for the self-similar thinning behavior modulated by the adsorbed BSA layer. Considering an infinitesimal element of length $l$ around the neck, the balance between $f_C$ and $f_{SV}$ gives $\gamma/R\sim\mu_sw_z/R$ and Eq. \ref{eq:S} yields $dR/dt \approx -Rw_z/2$. We thus have $dR/dt \sim -\gamma R/\mu_s$, where $\mu_{s}=\mu_{s0}\Gamma/\Gamma_0$. The conservation of BSA at the surface reads $\Gamma Rl=\Gamma_0 R_0l_0$ and the liquid volume conservation reads $R^2l=R_0^2l_0$, yielding $\Gamma =\Gamma_0 R/R_0$. We find
\begin{equation}
  r_{min} \sim \frac {1}{Oh_{s\mu}} \tau,
\label{eq:Scaling}
\end{equation}
which well describes the neck radius evolution near pinch-off at a large $C$ in Fig. \ref{fig:scalinglaw}.

In conclusion, we document that the satellite drop during the rupture of an inviscid liquid thread is fully suppressed with the presence of a protein adsorbed viscoelastic surface layer. Our theoretical model identifies the dominant role of surface shear viscosity at the late thinning stage, which inhibits inertia and the associated nonlinear effects, thus effectively keeping the fastest thinning position at the thread center. We show the balance of surface shear viscosity with capillarity yields a unique self-similar thinning behavior near pinch-off. This study exemplifies that surface rheology profoundly reshapes the dynamics of multiscale free surface flow in a way distinct from bulk effects and Marangoni effects. By only modifying the thread topology at a small spatiotemporal domain near pinch-off, surface rheology barely influences the global flow, manifesting a potential passive strategy without changing bulk rheological properties to eliminate satellite drops in practical scenarios of precise drop generation. Meanwhile, our findings may advance the understanding on the aerosol transmission via liquid fragmentation or spray in exhalation activity, bioengineering, and the food industry, where macromolecule-induced viscoelastic surface are ubiquitous.

\vspace{\baselineskip}
\emph{Acknowledgments} -This research was supported by the National Natural Science Foundation of China (Grant No.12272026, U2341281, 12502287), and the Natural Science Foundation of Beijing Municipality (Grant No. L248008).

\bibliographystyle{prsty_withtitle}
\bibliography{References}

\end{document}